# Impact of a small number of large bubbles on Covid-19 transmission within universities

Alan Dix

Computational Foundry, Swansea University, Wales, UK

## Abstract

This paper uses a variety of analytic and computational models to assess the impact of university student social/study bubbles.  Bubbles are being considered as a means to reduce the potential impact of Covid-19 spread within Universities, which may otherwise indirectly cause millions of additional cases in the wider population. The different models agree in broad terms that any breaking of small bubbles into larger units such as a year group or small student halls, will lead to substantial impact on the larger community.  This emphasises the need for students to be well-informed and for effective campus test, track and trace.

## Headlines

- without strong controls, the return to universities would cause a minimum of 50,000 deaths
- the impact will be felt principally in local communities rather than the students themselves
- student social/study bubbles have been suggested as a way to reduce this impact
- to be effective bubbles have to be small (approx a dozen) and very strictly maintained
- as a rule of thumb a bubble size of 10 will increase overall population R by 10-20%
- if bubbles 'leak' into wider groups of 50-100 this leads to larger scale outbreaks

## Actionable points

- it is critical that students understand how important it is to minimise transmission
- outbreaks are probably inevitable increasing the importance of internal track-and-trace
- minimising student–non-student transmission is also critical to protect communities

# 1. Introduction

In the autumn schools and universities across the UK will reopen welcoming around two million students back to campuses across the four nations.. This reopening is critical for social, educational, and economic reasons (e.g. see [WFA20]). However, it does carry risk.

The current UK central government Covid-19 strategy is to reduce the rate of transmission, but not eradication. Given this policy, large gatherings, especially over protracted periods, have the potential to create large, hard to control spikes. Universities fall into this category and without mitigating measures the autumn return could cause thousands of death and/or local or national lockdowns.

If no controls are exercised on campuses, then a rough but optimistic calculation based on the student body size and low figures of R means that the return to campus would cause an additional 50-60,000 deaths across the country, mostly amongst the elderly and vulnerable. In practice, outbreaks in universities would have even larger localised effects and lead to local lockdowns and/or quarantining of campuses. (see section 2.1)

Whether in terms of deaths or draconian measures, this scale of impact is clearly unacceptable, even given the clear benefits, and so universities are actively seeking ways to mitigate and control spread on campus and between campus and outside population.

Even this level of rough calculation offers ways of considering mitigations. These are effectively of two kinds:
1. Reducing spread within campus – this is the main focus of the modelling within the rest of this working paper.
2. Limiting the spread into wider society, effectively treating the university as a large self-quarantine unit.

Both of these are important and interact.

Universities are establishing guidelines to minimise spread within the university body including increasing hygiene, greater use of online materials and social distancing within any col-located interactions [UUK20a].

However, students normally have far closer day-to-day social interaction than general society including shared accommodation, and many social activities. This is known to give rise to infections notably the regular phenomenon of 'freshers' flu' as well as more occasional meningitis outbreaks. One suggestion to limit the impact of this is to encourage students to form social and educational bubbles, whereby groups of students on the same course are housed with one another in student accommodation and attend the same tutorial groups etc., thus limiting outbreaks to the bubble in the same way that lockdown has for ordinary households.

However, even if bubbles are effective at limiting direct between-student spread, indirect spread through the wider population may 'break' social/study bubbles leading to runaway infection. This emphasizes the dual interacting role of both kinds of mitigation above.

These student bubbles have some similarities with super-spreaders and super-spreading events, which can have major short-term and localised impact on Covid-19 control and are well studied.  There are also studies of the impact of widespread small bubbles (e.g. paired households) [LW20].  However, this working paper considers relatively small numbers (1 or 2% of population) of relatively large bubbles (10 or more).  Figure 1 places this work in relation to other scales of bubble-modelling.

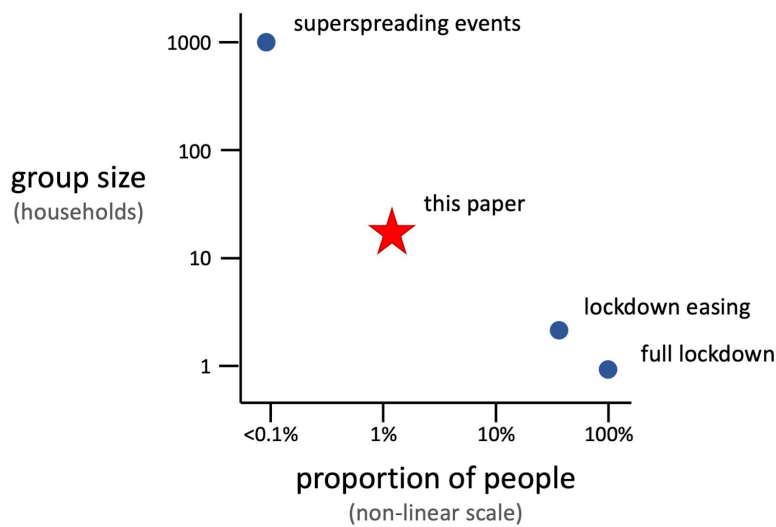

**Figure 1.  Orders of magnitude of bubble sizes and frequency**

The next section presents the initial order-of-magnitude estimates of the impact of unrestrained campus re-opening (socially distanced, but not 'bubbled') and also discusses in informal terms the two flows of infection within the student body and to the wider population.  This is then followed by detailed analytic modelling of large bubbles in Section 3 and computational modelling in Section 4.

Section 2 will refer forward to the detailed modelling sections, but is presented first in order to give a qualitative understanding of the issues feeding into the models and also to offer an overview for the non mathematical reader.

Note that section 3 and 4 deliberately choose the simplest model possible to study the relevant effects.  This makes it easier to avoid errors (easy to make in this kind of work) and also easier to comprehend the broad processes and hence easier to look at potential mitigation and control strategies.

In addition, I hope that these models will be suitable raw material for exercises for mathematics or computer science students.  This is in the hope that by understanding the reasons for different control measures on campus and the dangers if these are flouted, these students can become advocates amongst their peers.

# 2. Initial discussion

## 2.1 Rough impact estimate

To gauge a round figure for the potential impact of University returns, let's assume all efforts to limit spread within campuses fail.  The current (July 2020) infection rate in the UK is around 1 in 2,300 [ONS20b]; although this will reduce somewhat by autumn 2020, this still means that every campus will be 'seeded' by at least some infections .  Even a relatively small within-campus  R of 3 (the approximate pre-lockdown figure for the UK), would effectively mean that a single seed will saturate the campus over a period of a term (R=3, inter-infection period ~3.5 days => ~ 10 times increase per week).

In the UK in 2018-2019, there were 2.38 million students of which 1.9 million were UK students [UUK20b].  If, as is hoped, most of the UK students do attend university in the coming year, then this accounts for around 2.5% of the UK population.  These effectively form a seed infection for the rest of the population, just as winter flu season hits and wet weather makes social distancing harder.  The student body itself would be expected to have relatively low health impacts and mortality especially if vulnerable students choose to study remotely.  The main impact will be on the wider population (see also [ULHES]).

It is arguably optimistic to believe that social distancing plus track and trace can keep the general population R below 1 during this period, but optimistically we could look at general population R figures of between 0.7 and 0.9, as was achieved during lockdown.  Crucially, spread within the population means that initial seed cases led to *cumulative consequences* in both the short and long term.

On the *most optimistic* side, an R of 0.7 means that each seed case leads to an additional 2.3 cases on average (1/(1-R)-1), that is between 5% and 6% of the general population will become infected due to the university impact.  This is as many as during the whole of the 'first wave' of Covid-19 in the UK, which the NSO estimate to have caused approximately 50,000 deaths at 1st May [BBC20] and approximately 60,000 by the end of June [ONS20].

On the *less optimistic* (but far from pessimistic) side an R of 0.9 means 9 additional cases for each seed case, that is 22.5% of the population eventually infected due to the student body, with around half of these within the winter flu season. If unaddressed this would clearly overwhelm the health system and lead to mass fatalities. In practice the impact would be first seen locally and lead to localised lockdowns of university towns, so the eventual mortality may well be no larger than the most optimistic case, but with widespread effects on the economy and society.

If, as is possible during the winter period, the general population R rises above 1, then there will be effectively slower or faster exponential growth in society anyway. This may be mitigated by increased efforts to track and trace or local semi-lockdown measures. In such a scenario, additional seeding on the scale that universities pose, will exacerbate the situation and may render less draconian measures ineffective.

## 2.2  Reducing within-campus spread

One of the ways within-campus spread may be managed is by grouping students into social bubbles with relatively large internal contact, but reduced contact between student bubbles. In addition, there are also a variety of legitimate and illegitimate activities that have a similar effect of bringing a relatively small proportion of the population together into 'super-bubbles' of tens or larger. This includes certain kinds of factories where social distancing is hard, pub-lock-ins, and groups of young people congregating informally.

This paper uses a variety of simple analytic and computational models to explore the impact of small numbers of large bubbles. Informally the immediate effect of the large bubble is to rapidly increase the rate of infection *within* the bubble, but of course those effects soon leak out. Effectively a bubble of size n has the impact of increasing the societal infection rate in a somewhat similar way to a super-spreader with n times as many contacts as normal.

The analytic modelling in sections 3 will use this analogy with super-spreaders and techniques of two-timing used in nonlinear differential equations to look at early spread (when the number of cases is low) and how bubble size impacts the effective societal R factor.

Given the use of multi-household bubbles as a way to open-up strict lockdown in a controlled fashion, there has been considerable interest in the impacts of this including mathematical modelling [LW20]..

However, the modelling of multi-household bubbles assumes a relatively high proportion of the population in relatively small bubbles. In contrast, for universities we are thinking about a relatively small (but not insubstantial) proportion of the population in relatively large bubbles.

The size of the university bubble is critical. Some suggestions are of bubbles of around a dozen students that are housed together on campus and share common tutorial groups, lectures etc., so that within bubble spread is expected to be extremely high, but between bubble spread is controlled.

The computational modelling will focus on order-of-magnitude bubble sizes of 10, 100, and 1000 corresponding roughly to the proposed 'social and study' bubbles, a typical year group within a course, and the overall students within a broader subject area or building on campus. The next order of magnitude 10,000 is effectively unconstrained growth and dealt with adequately by the rough calculations in section 2.1.

As well as helping to control infection, bubbles will help efforts to track and trace once cases are discovered. This will be especially important amongst students who are more likely to be asymptomatic. This and other forms of additional measures are not modelled in detail, but the impact of larger bubbles, cross-bubble contamination is substantial which emphasises the importance of such measures.

## 2.3  Limiting the spread between campus and wider society

Most of the modelling in this working paper will assume that spread between students and other members of the population will be at the 'normal' population level of contact. This might be an overestimate of cross-infection (especially for self-contained out-of-town campus universities) as students are largely interacting with one another, or might be an under-estimate (especially for city or multi-campus universities) as students will need to travel a lot on public transport between halls and campuses, eat out, and maybe be more 'risky' in terms of night-life, etc.

However this potential cross contamination is crucial for two reasons.
1. The major impact in terms of fatalities is likely to be in the general population, both because of amplification effects of any non-zero population R, and because the mortality rate is far higher in the broader population (~1%) than students (<0.1%).
2. Feedback effects back-and-forth between students and wider society amplify infection rates.

The former is well understood since early figures of differential risk emerged from Wuhan and have been confirmed by further studies and are evident in university modelling both within this paper and elsewhere [ULHES].

Feedback effects can be harder to understand and Figure 2 summarises the main flows between student body and the wider population: (i) the rapid growth within the student body leads to (ii) seeding of the wider population and then (iii) cumulative growth given any non-zero R; the higher population incidence then (iv) causes cross infection back into the student body.

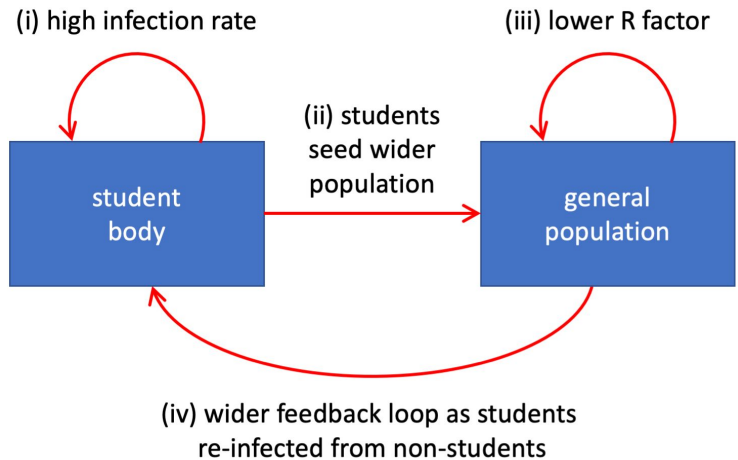

**Figure 2.  Feedback effects between students and wider population**

In general the rates of growth in a dynamical system are driven by the most powerful feedback loop.  In the worst cases of intra-student transmission, the internal feedback (i) is so rapid the external effect is negligible.  However, we will see (section 4.3)  that where student bubbles restrict growth on campus, the external feedback loop effectively breaks the bubbles from the outside and drives growth.

## 2.4  Locality and disproportionate impact

The modelling here will assume perfect mixing outside campus, as in everyone in the population is equally likely to infect any other person.  However, in reality there are strong geographic effects and university students are a much larger proportion of the population of the areas in which they reside.

For example there are nearly 100,000 university students in Manchester compared with a population of 500,000, that is nearly 20% as opposed to 2.5% national figure, in Aberystwyth students are ⅓ of the term time population and in St Andrews students outnumber the rest of the population of the town.

This locality effect could be positive at a national scale as any runaway outbreaks can be detected and contained more rapidly.  Furthermore it reduces cross-seeding between universities.  However, the impact on the communities within which they reside will be proportionately greater, both in terms of potential infection risk and the likelihood of requiring emergency lockdowns.

In addition, however there are negative consequences beyond the locality.  The greater the proportion of the student population, the stronger the student–population feedback effect (Figure 2.iv) leading to faster overall growth in numbers infected.

Crucially urban and lower-tariff universities may tend to have a higher proportion of live-at-home students in communities already likely to be disproportionately affected by Covid-19 (higher social deprivation, higher BAME proportion).  It will be far harder to control student-population spread in such circumstances.

# 3. Analytic models of early spread

The author's first introduction to the impact of large bubbles was driving through a village just before lockdown and noticing approximately 20 young people gathered, not unsurprisingly without appropriate social distance.  The question that came to mind was how significant this kind of grouping would be.  Clearly if one of the young people were infected, it would rapidly spread within them as a group and then to their families, but if this teen-age super-bubble is still a relatively small proportion of the overall village population, how much effect does that have on the community as a whole?

As noted earlier, one way to think of these groups is rather like super spreaders.  Treat the group as a single supra-individual (if one is infected they are all infected) but with n times the number of contacts as an average person.

Note that this analogy is a form of 'two timing' method as used in nonlinear differential equations.  We have a fast time scale when the virus spreads within the group and then a slower timescale at which within-group spread is considered instantaneous and in which we model wider population growth.  Given transmission within a group is not really instantaneous, we have to do some later adjustments to interpret the results.

We'll first look at simple individual super-spreaders, then the group-scale modelling, and finally the adjustment process.

## 3.1 Super-spreaders

In the early days of the Covid-19 outbreak, there was considerable discussion of super-spreaders and also super-spreading events.  Both of these vastly increase the variability of infection, increase the likelihood of localised or short-term overload of the health service and make track and trace very difficult.  However, they are relatively easy to control and to some extent 'average out' if they are sufficiency infrequent in society.

In the early spread of a disease the number of infected people is driven by a simple linear equation giving rise to linear growth:

Covid-19 – Impact of a small number of large bubbles on University return    9$$dI/dt = \gamma\, I$$

so

$$I = \exp(\gamma\, t)$$

where $\gamma$ is the new infection rate ($\alpha$) minus the recovery rate ($\beta$).

Given an initial seed infection, the total number of new infections (R) due to this is:

$$R = \alpha \int_0^\infty \exp(-\beta\, t)$$
$$= \alpha / \beta$$

Let's say that a proportion p of the population are superspreaders and that these infect k times as much as the general population ( $k\,\alpha$ vs $\alpha$ ).

So the total newly infected people are:

$$d\, I\_new / dt = \alpha\, (\, I\_pop + k\, I\_ss\, )$$

Of these 1-p will be normal population and p of them superspreaders, so:

$$d\, I\_pop / dt = (1-p)\, d\, I\_total / dt \; - \; \beta\, I\_pop$$
$$d\, I\_ss / dt \;\;\; = p\, d\, I\_total / dt \; - \; \beta\, I\_ss$$

Together we get the simple vector equation:

$$d\,(I\_pop,\, I\_ss)/dt \;=\; (1-p)\,\alpha \begin{bmatrix} (1-\mu) & k \\ \varepsilon & \varepsilon k - \mu \end{bmatrix} (I\_pop,\, I\_ss)$$

where $\varepsilon = p/(1-p)$
and $\mu = \beta / ((1-p)\alpha) \sim 1/R$ (note if $\mu > 1$ the R < 1 and the infecion dies)

Solving for eigenvalues of the matrix ($\lambda$), we get:
$$0 = (1-\mu-\lambda)(\varepsilon k - \mu - \lambda) - (1-\mu)(\varepsilon k - \mu)$$
$$= \lambda^2 - (1+\varepsilon k - 2\mu)\,\lambda$$

So
$$\lambda = -\mu \;\; \text{or} \;\; \lambda = 1 + \varepsilon k - \mu$$

The larger value is the major growth and when multiplied by $(1-p)\,\alpha$ gives us an overall growth rate of $((1-p) + pk)\,\alpha - \beta$. recalling that $R = \alpha/\beta$ , that is the overall growth rate is exactly the weighted average of the different growth rates of individuals.

## 3.2 Large bubbles

We'll assume that a small proportion p of the population are in bubbles of size n.   Taking a two timing model, we assume that if any member of a bubble is infected, then the rest are instantaneously infected also.

We can translate this directly into a superspreader model, by treating each bubble as a supra-individual who spreads at a rate n times faster (as there are n people in the group). There are only 1/n of these groups, so the effective p is also smaller by a factor of 1/n.  Given the critical factor was pk, this initially looks like the bubbles have an insubstantial effect. However, the analogy is not perfect as (critically) a bubble is n times as likely to be infected as a single individual.

Instead, we can use a similar derivation for the superspreader, with each individual separately, but take into account group spread as a separate step.

The total newly infected people are:

$$d\,I\_new\,/\,dt\ =\ \alpha\,(\,I\_pop + I\_bub\,)$$

Note that here we assume individuals in a bubble are no more or less risky outside their bubble, so they have no additional spread.

As before (1-p) of these are in the general population::

$$d\,I\_pop\,/\,dt\ =\ (1\text{-}p)\,d\,I\_new\,/\,dt\ -\ \beta\,I\_pop$$

However, for each initially infected person in a bubble immediately the whole bubble of n is infected:

$$d\,I\_bub\,/\,dt\ =\ n\,p\,d\,I\_new\,/\,dt\ -\ \beta\,I\_bub$$

The simple vector differential equation now becomes:

$$d\,(\,I\_pop,\,I\_bub\,)\,/\,dt\ =\ (1\text{-}p)\alpha\begin{bmatrix}(1-\mu) & 1 \\ \varepsilon n & \varepsilon n - \mu\end{bmatrix}(\,I\_pop,\,I\_bub\,)$$

where  $\varepsilon = p/(1-p)$  as previously
and  $\mu = \beta / ((1-p)\alpha)$

Although this is slightly different from the superspreader equation, it is essentially the same and so again, solving for the eigenvalues ($\lambda$) of the matrix, we again get:

$$\lambda = -\mu \quad \text{or} \quad \lambda = 1 + \varepsilon n - \mu$$

That is an overall growth rate ((1-p) + pn) times larger than it would otherwise have been. Given p is small and n is large, this is approximately (1 + pn).

## 3.3 Impact on R value

We saw that both superspreaders and large bubbles increase the effective R value by a proportion that is the weighted average of their frequency and the underlying population growth rate.

For the superspreaders these equations are fairly accurate. The real process is discrete and probabilistic, rather than driven by real-valued differential equations. In the very earliest stages of spread, or for effective track and trace when levels of infection have been reduced sufficiently, the stochastic nature of infection and in particular the additional variation driven by superspreaders are very important, and so different analytic methods are used. Also at later stages of epidemic spread, when there are significant numbers of resistant individuals, the equations need to be modified, for example with a SIR model, which we will use later in section 4. However, for mid-early stages the differential equations are likely to be a good fit.

For large bubbles however, we need to perform some adjustments. While the additional contacts of a superspreader will happen within the normal period of infection (before the carrier becomes symptomatic or recovers). In contrast, the two-timing approach, which assumes within-bubble infection is total and instant, is a more substantial approximation.

For classic uses of two timing the additional terms that are modelled by the slow change are over many timescales of the faster phenomenon, for example, models of precision of orbits. However, here the spread within bubbles is faster than within the population, but not so much faster.

Thinking of the order of magnitudes for the university case: 10, 100, 1000. It is reasonable to assume that spread within the immediate socal bubble happens within a single cycle of infection, but spread to 100, 1000 probably take 2 and 3 cycles respectively.

Thinking about the first subcase. Over two cycles of infection (approximately a week) the single bubbled spreader could be treated as 'instantly' becoming 10 times the start of week infection, but the effective 'R' this should be compared with is the number of people infected over two cycles, that is R^2. That is instead of a multiplier of (1+np), it would be $\sqrt{1+np}$. Similarly, if we assume the first order bubbels break down and infect the next (100), we need to cube root 1+np, and so on.

Table 1 shows this tabulated for the three orders of magnitude (r) showing the overall increase of the population R.  Table 2 does similar calculations, but adds two cycles in each case, that is 'averaging out' the fast growth for 1 order of magnitude over 3 cycles, second order of magnitude over 4 cycles and the third over 5 cycles.   Finally Table 3 takes an even more concervatve approach for the larger bubble sizes and averages out the within-bubble growth over twice the expected bubble saturation time, that is 2, 4 and 6 cycles respectively.

**Table 1 – proportionate increase in population R (1 cycle added) [MD1]**

| r | g^r | 1+np | R' | increase |
|---|-----|------|----|---------| 
| 1 | 10 | 1.2 | 1.09544512 | 10% |
| 2 | 100 | 3 | 1.44224957 | 44% |
| 3 | 1000 | 21 | 2.14069514 | 114% |

*Assumes 2% of population in bubbles, and*
*r= order of magnitude, g^r is maximum bubble size, and*
*R' = proportionate change in overall population R value*
*This calculation is based on adding 1 cycle to within-bubble saturation time*

**Table 2 – proportionate increase in population R (2 cycles added) [MD1]**

| r | g^r | 1+np | R' | increase |
|---|-----|------|----|---------| 
| 1 | 10 | 1.2 | 1.06265857 | 6% |
| 2 | 100 | 3 | 1.31607401 | 32% |
| 3 | 1000 | 21 | 1.83841629 | 84% |

*Assumes 2% of population in bubbles, and*
*r= order of magnitude, g^r is maximum bubble size, and*
*R' = proportionate change in overall population R value*
*This calculation is based on adding 2 cycles to within-bubble saturation time*

**Table 3 – proportionate increase in population R (doubling cycles) [MD1]**

| r | g^r | 1+np | R' | increase |
|---|-----|------|----|---------| 
| 1 | 10 | 1.2 | 1.09544512 | 10% |
| 2 | 100 | 3 | 1.31607401 | 32% |
| 3 | 1000 | 21 | 1.66100096 | 66% |

*Assumes 2% of population in bubbles, and*
*r= order of magnitude, g^r is maximum bubble size, and*
*R' = proportionate change in overall population R value*
*This calculation is based on double the within-bubble saturation time*

First note that the final figure in each table is the impact on the overall population R value, not just the rate of growth within the student population.

For bubbles of size 10 the impact on the overall population R factor is between 6 and 10% depending on the more or less concervative assumptions.  These feel manageable, but of course not insubstantial if R is close to 1.  However, if the small bubbles break and the effective bubble size becomes 100 (subject year group), then we are looking at societal increase of R of around 30%.  That is if R would otherwise be 0.7, it will end up close to 1, and if it is 0.9, then the student body will tip it to an R of 1.2, which is quite fast societal growth (doubling every two weeks).  The third order of magnitude is patently calamitous, multiplying the societal R factor by somewhere between an additional 50% and double its value without the student effect.

# 4. Computational models

In this section we consider two models.

The first (Section 4.1)  uses discrete time dual SIR model (normal population and bubble groups) using two timing approximations.

The second (Section 4.2) uses a multi-generational SIR model for the individuals in groups, which means it does not need to use two-timing approximations and hence a simpler model conceptually, albeit harder to model analytically.

## 4.1 Dual population SIR model with two timing

For live data see:
https://drive.google.com/file/d/1vCDc-p3DtInyZBoZxHub_nyKcHaSWUF8/view?usp=sharing
Note for this link may change, for permanent reference use [MD2].

This model uses two SIR populations.  One corresponds to individuals in wider society, the other are groups as supra-individuals.

A normal (discrete) SIR model would be:

$$C\_new = \alpha\ I \qquad \textit{fresh contaminations}$$
$$\Delta S = -C\_new * (S / P)$$
$$\Delta I = C\_new * (S / P) - \beta\ I$$
$$\Delta R = \beta\ I$$

Where $\alpha$ is the infection rate, $\beta$ is the recovery rate and P the overall population size. For for the simulations reported here $\beta = 1$ is always used, which effectively means everyone recovers after a single cycle of modelling, a simplification that seems common in the literature as people either recover or are symptomatic and isolate relatively rapidly. In this case $\alpha$ is the R factor for the normal population.

For the two cohorts we modify this to have two populations ( Sp, Ip, Rp ) for the normal population and ( Sg, Ig, Rg ) for the groups in bubbles. In this model a group is treated as a single supra-individual and so both infects n times as many people (where n is the bubble size) and is also infected n times as often (if any individual is infected, the whole group is).

That is this is a two timing model where we assume infection within the bubble is effectively instantaneous. This needs correcting (as we did in section 3.3 for the analytic models), to turn the modeling results into ones that relate to real world times and R values.

The two cohort discrete SIR model is therefore:

$$C\_new = -\alpha\ (Ip + n\ Ig) \qquad \textit{fresh contaminations}$$

$$\Delta Sp = -C\_new * (Sp / P)$$
$$\Delta Ip = C\_new * (Sp / P) - \beta\ Ip$$
$$\Delta Rp = \beta\ Ip$$

$$\Delta Sg = -C\_new * n\ (Sg / P)$$
$$\Delta Ig = C\_new * n\ (Sg / P) - \beta\ Ig$$
$$\Delta Rg = \beta\ Ig$$

The model is run over 30 cycles of infection, which, taking into account two timing, is approximately a year (1 week ~ 1 cycle) for n=10 and approximately two years for n=1000.

The model is run over a range of population R values from 0.7 to 3 and bubble sizes from 10,to 1000 and proportions of population in large bubbles between 0.5% and 10%. The ranges are broad to validate the robustness of the analytical early growth model in section 3.2. However even some of the unrealistic values, notably 10% for proportion in the large bubbles, would correspond to local conditions if not nationally..

Over all parameter values the initial effective population R value matches the analytic model to at least 4 digits (see tables in the results sheets on [MD2]).  That is the effective population R is magnified by a factor of (1–(n–1)p) where p is the proportion in bubbles and n is the bubble size.

Looking in greater detail, while the multiplier changes continuously with the group size, in fact many of the overall impacts over the lifetime of the model saturate once the bubble size is around 50 (varying with R and %in bubbles).

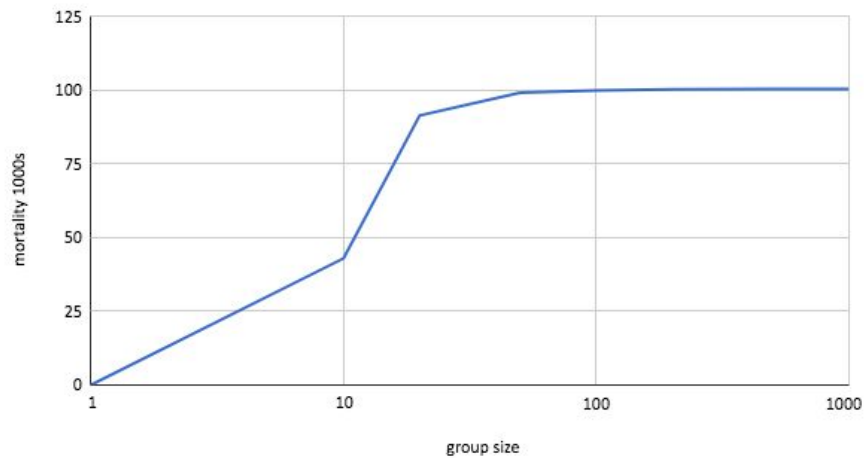

**Figure 3.  Saturation of overall mortality with different bubble size
(R_pop = 1.0 and p = 2%)**

This saturation is largely because the population in groups has been hit peak infection before the end of the simulation.  See for example, Figure 4 for a small group size (n=10) where the infections are still growing linearly after 30 infection cycles (~ 6 months, an academic year), whereas in Figure 5 with a medium sized bubble (n=100) the bubbled population has been completely infected with virtually no new cases by cycle 10 (probably about 15-20 weeks given two timing adjustment) even with s smaller normal R factor (0.9).

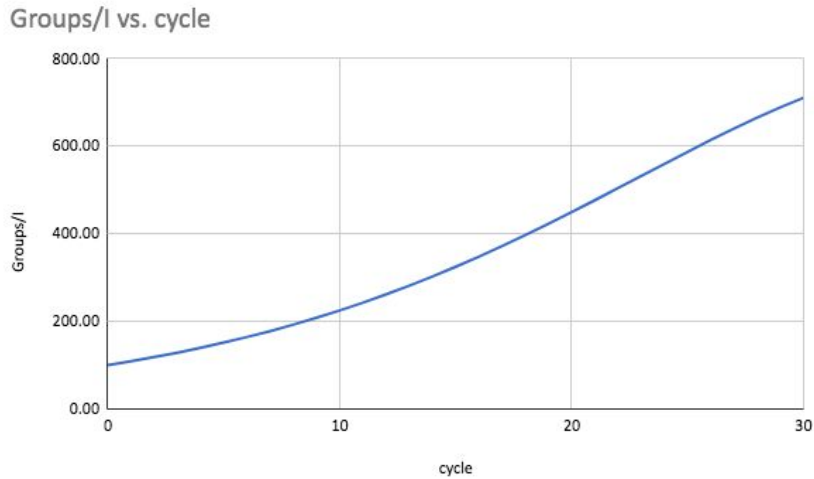

**Figure 4.  Number of new student infections with R_pop = 1.0 and group size of 10**

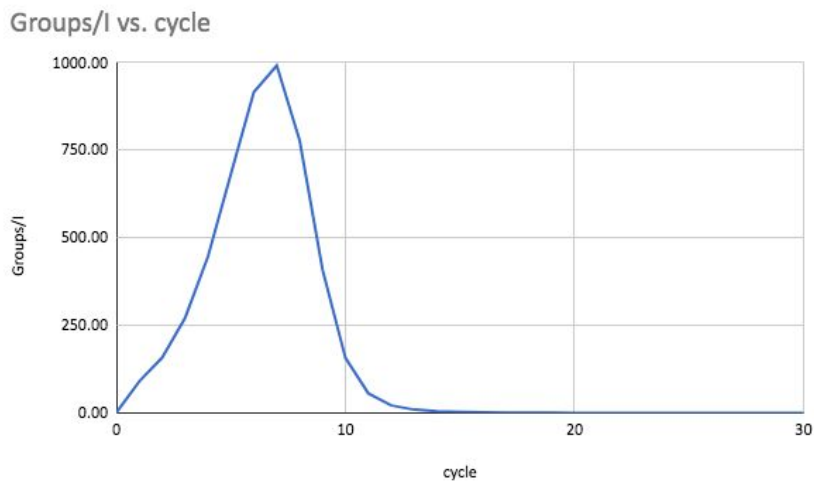

**Figure 5.  Number of new student infections with R_pop = 0.9 and group size of 100**

For the cases where there is saturation, the very rough calculations in Section 2.1 would be sufficient to estimate the overall impact.

These simulations emphasise how critical it is to maintain the smaller bubbles.  If there is even relatively small breaking of these, the impact on societal mortality is disproportionately large.

Finally, it is worth noting the impact of breaking small bubbles is, not surprisingly, most serious when the societal R value is otherwise close to 1.  Figure 6 shows excess mortality in the wider population (excluding students) for bubble size of 100 (class/year group) for different values of R.  For larger R the disease is effectively epidemic anyway, so the impact of the universities is inconsequential.  For smaller R, the main infection is within the student body (who are likely to have low mortality rates) and the rest of society relatively untouched (although there may be local effects as noted in section 2.4).

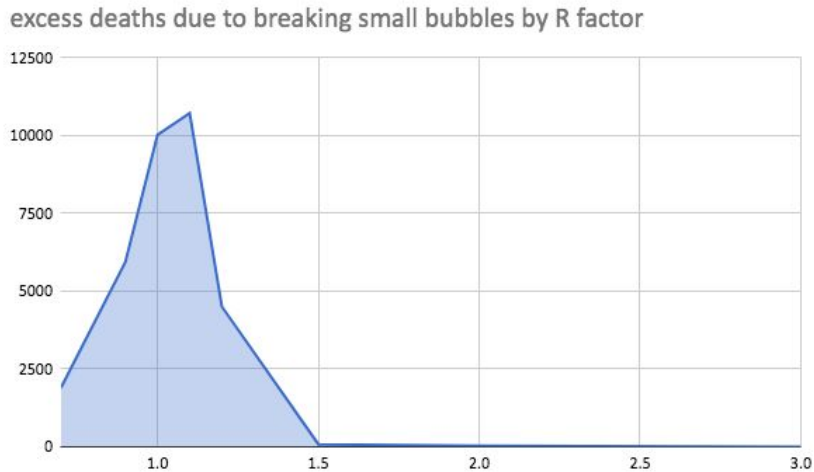

**Figure 6.  Excess mortality to medium university bubbles (n=100,p=1%),
            showing sensitivity for R near 1.**

## 4.2 Dual-cohort generational model (no two timing)

For live data see:
https://docs.google.com/spreadsheets/d/1_GcLVkdsY3V3pY_LVWinWL5XXktXkWZ1CD148tsu-s8/edit?usp=sharing
Note for this link may change, for permanent reference use [MD3].

The two timing makes easier models, but adds some complexity when converting back to 'normal time.   We will therefore validate this by using a model that is simpler to model computationally and closer to real transmissions.

We again use a two cohort discrete SIR model with (Sp, Ip, Rp) for the general population, but multiple infection generations for the individuals in bubbled groups (Sg, I1, I2, I3, I4, Rg).  I1 represents the individuals who have been infected by general population transfer, including from individuals in other bubbles.  I2 are those in the bubbles who have caught the disease from those in I1, etc.

We use a recovery rate ($\beta$) of 1 throughout.

$$C\_new = -\alpha\,(Ip + I_1 + I_2 + I_3 + I_4)  \quad \textit{normal population contamination}$$

$$\Delta Sp = -C\_new * (Sp / P)$$
$$\Delta Ip = C\_new * (Sp / P) - Ip$$
$$\Delta Rp = Ip$$

$$\Delta Sg = -C\_new * n (Sg / P)$$
$$\Delta I1 = C\_new * n (Sg / P) - I1$$
$$\Delta I2 = n_1 I1 - I2$$
$$\Delta I3 = n_2 I2 - I3$$
$$\Delta I4 = n_3 I3 - I4$$
$$\Delta Rg = I1 + I2 + I3 + I4$$

The model is not precisely comparable with the two-timing model as the time steps in this model correspond to normal infection time of the disease (approx ½ week) rather than stretched times and the R values are real R0.

However, the qualitative behaviour is similar, as can be seen from figures 7–10, which parallel figures 3–6.

First in figure 7, we see that the excess mortality due to different bubble sizes grows and then saturates rapidly. That is, if the initial bubble size is 10, then even a small number of 'broken' bubbles is enough to make dramatic differences in societal mortality. By the time this has reached typical subject year-group size (60-200 students), the impact is nearly as bad as complete runaway infection across campus.

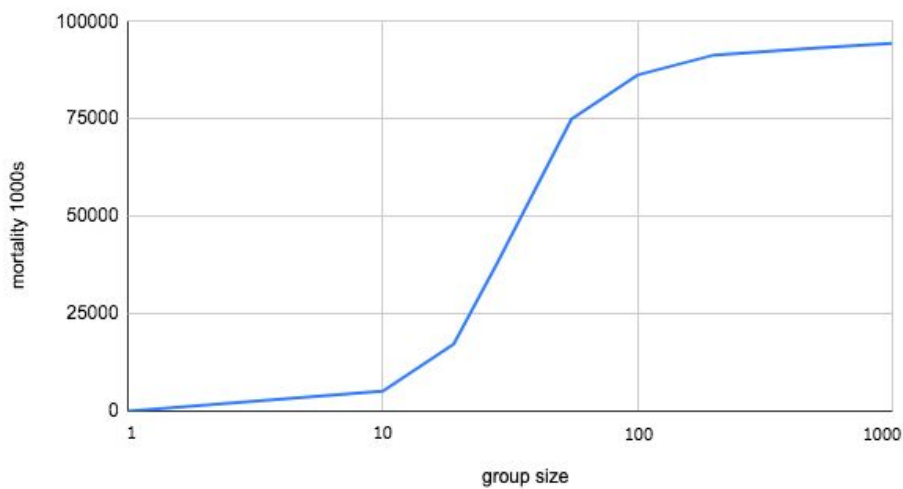

**Figure 7. Generational model – Saturation of overall mortality with different bubble size (R_pop = 1.0 and p = 2%)**

The reason for this dramatic effect is evident if we look at new student infections. Just as in the two timing model in section 4.2, student infections continue to grow throughout the year when the bubble size is 10 (figure 8). This is increasing the societal R value, but in a relatively constrained manner. In sharp contrast a bubble size of 100 (figure 9), even with a lower initial R value, leads to an uncontrolled 'wave' which effectively means the entire student body has been infected by the end of the first term. Even if there is no direct student-student transmission between bubbles, the town–gown feedback loop in figure 4 can lead to sufficient bubble–bubble transmission given the large within-bubble magnification..

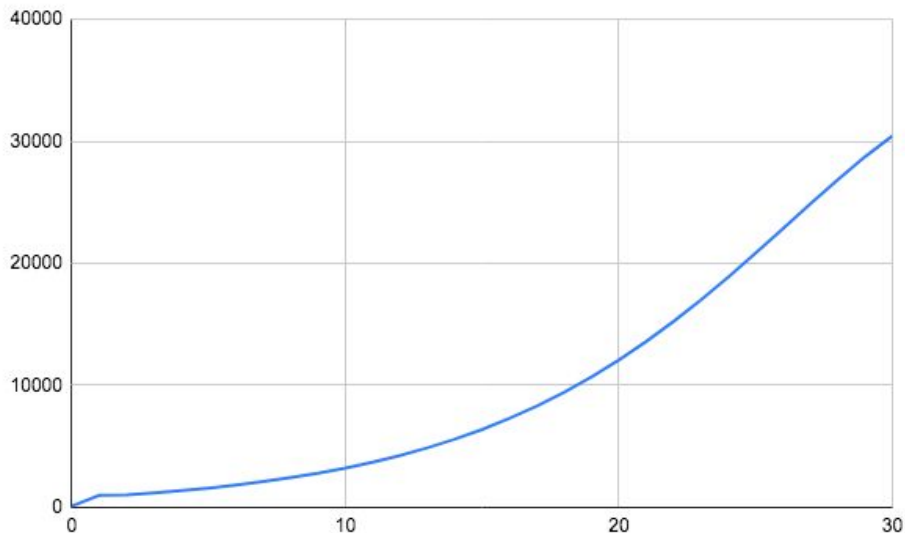

**Figure 8. Number of new student infections with R_pop = 1.0 and group size of 10**

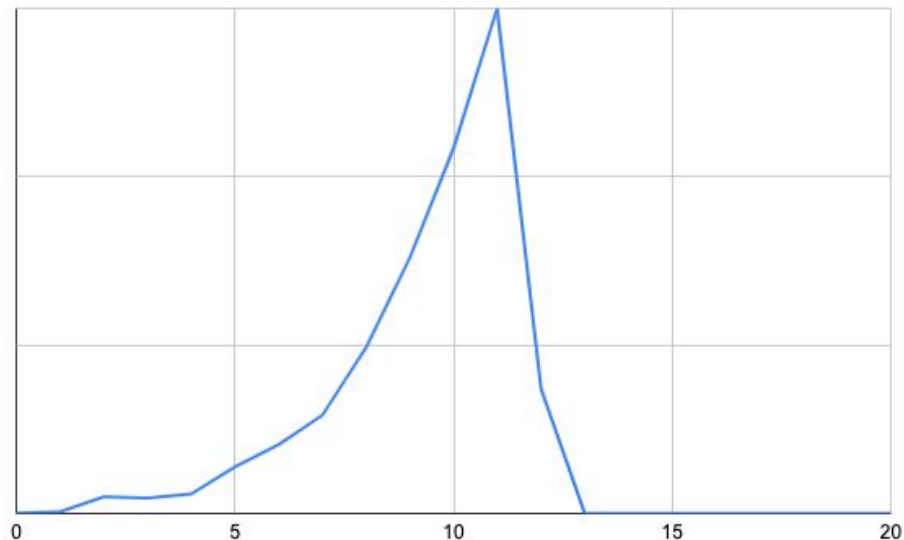

**Figure 9.  Generational model – Number of new student infections with R_pop = 0.9 and group size of 100 (p=2%)**

We also see that, like the two-timing model, the impact of larger student bubbles is most dramatic for R values near 1.  For larger values of R there is effectively runaway societal infection anyway.

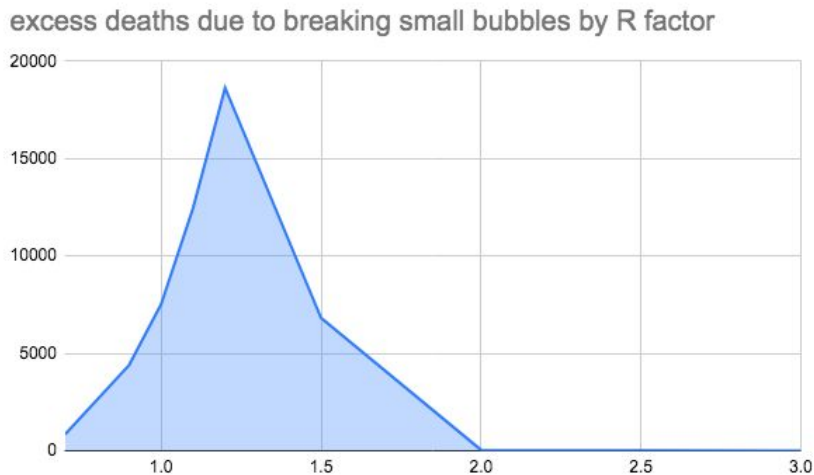

**Figure 10.  Generational model – Excess mortality to medium university bubbles (n=100,p=1%), showing sensitivity for R near 1.**

As in the discussion of previous models the large mortality figures would be unlikely to materialise as there would be local or campus lockdowns, but effectively they indicate situations where the student body is causing substantial impact on the surrounding community.

## 4.3 Unforeseen modelling outcomes – societal cumulation and freshers' flu

While the generational model validated the qualitative features of the two-timing model, there were two surprises, both of which included effects that were larger than expected.

1. Larger and longer lasting impact outside the student body
2. Larger than expected effective population-wide R factor

On the first point, one might imagine that the student body would become a fast-growing sub-population that then seeds the wider population.  However, the proportion of cases outside the student body grew faster than this would initially suggest.  The reason for this is that unless R is very small, the impact of 'seed' infections from the student community is cumulative – each initially infected person infects R others, etc until there are eventually (1/(1-R)-1) additional population cases due to a single infected student.  This accumulating population case load then re-infects other bubbles.  This (in retrospect obvious!) insight allows the robust back-of-the-envelope style calculations of section 2.1.

On the second point, recall that the analytic modelling suggests that the bubbled student population will increase the effective population R by a factor of 1+np where n is the bubble size and p the proportion of the population in bubbles (~2% for university students).  However, this model uses a two-timing approach that is looking at 'slow' time where each unit of model time is several disease infection cycles.  In section 3.3 this slower time is taken into account leading to a final figure of $\sqrt[n+1]{1+np}$, still catastrophically large for bubble sizes greater than a dozen or so.

However, the generational model is using real time units where 1 unit = 1 cycle of viral infection. The effective R value should therefore be closer to this corrected value, but, in fact, for small np is virtually identical to 1+np.  This is substantially larger than expected.  However, this large value of Reff decays over time and is only at the unexpectedly high rate near the beginning of the infection.

Closer examination of the model step-by-step outputs shows that this is due to the fact that initially infected students will tend to be uniformly spread amongst the bubbles and the high transmission *within* the bubbles means a near instant multiplication of the infected number of students by the bubble size.  This is effectively the Covid-19 equivalent of fresher's flu.  Note that this early surge is based solely on multiplication within social/study bubbles and doesn't take into account  additional factors such as first year students' natural tendency to grow social contacts in the first few weeks of term.

This Covid-19 freshers' flu-like surge suggests that on campus testing and track and trace will need to be established very rapidly.

# 5.  Summary and Further Work

There is a real potential for the return of students to Universities to cause substantial spikes in Covid-19 cases leading to additional deaths and local or even national lockdown.  One suggested solution is the assignment of students to social/study bubbles of around 10 students encompassing all face-to-face accommodation, socialisation and learning activities.  This paper has examined the sensitivity of this if bubble boundaries are breached leading to larger effective bubbles, say throughout the year group within a subject area.  Analytic modelling and two different forms of computational model all suggest that the impact of larger bubbles is profound. Emphasizing both the importance of strict adherence and also mitigating measures such as extensive within–university testing.

The models used are relatively simple, but for that reason likely to be robust in terms of their qualitative behaviour.  However, there is clearly a need for more detailed modeling for a number of reasons:.

1. exploration of sporadic features –  The models used all use average infection rates, but it is also likely that the bubbles will increase variation, that is lead to sporadic outbreaks. These long tail effects (small numbers of large events) may be harder to control and lead to greater risk [CT20]
2. geographic locality – The models are based on uniform mixing outside of the bubbles, whereas real student populations are often a larger than average proportion of the local community (as high as 50% in St Andrews, see section 2.4).
3. Intervention modelling – Understanding the phenomena can help suggest fruitful intervention strategies, but more detailed modelling could help model the impact of specific interventions such as track-and-trace, especially in the presence of the above effects (1&2).

# Models and data

[MD1]  Correcting two-timing effects. *Covid-19 related working papers.*
http://alandix.com/academic/papers/Covid-WPs-2020/
[MD2]  Sensitivity to rapid growth of subgroups. *Covid-19 related working papers.*
http://alandix.com/academic/papers/Covid-WPs-2020/

[MD3]  Covid grouping cohort model.. *Covid-19 related working papers.*
http://alandix.com/academic/papers/Covid-WPs-2020/